\documentclass[conference]{IEEEtran}
\usepackage[utf8]{inputenc}
\usepackage[T1]{fontenc}
\usepackage{mathptmx}
\usepackage{amsmath,amssymb}
\usepackage{booktabs}
\usepackage{graphicx}
\usepackage{xcolor}
\usepackage{url}
\usepackage[hidelinks]{hyperref}

\title{The Price of Random Access:\\
Measuring Block Granularity Across Four Compressed Formats}

\author{\IEEEauthorblockN{Yakiv Shavidze}
\IEEEauthorblockA{Independent Researcher\\
ORCID: 0009-0008-3622-3448\\
\url{https://github.com/yasha1971-coder/aceapex}}}

\begin{document}
\maketitle

\begin{abstract}
Random access into compressed data is normally bought with density. We measure
the exchange rate. Across four formats and nine axes on a common corpus, the
cost of cutting a 254\,MB archive into independently addressable 16\,KiB units
is 1.632\% of the archive for an absolute-offset format against 6.57\% for
seekable zstd, and the gap widens as the unit shrinks: at 4\,KiB, 5.33\%
against 10.06\%. Because the cost is small, several properties follow that are
usually unavailable: splitting an archive is free and occasionally profitable
($-0.28\%$ on tiled input), append needs no format change, seek latency does
not depend on position, and one archive is read by both a CPU and a GPU
decoder. We give three structural results with proofs and bit-perfect
verification---that the repeat-distance chain of an LZ77 parse forms a
substitution monoid and is therefore prefix-scannable without touching the
bitstream, that self-overlapping matches are periodic rather than chained, and
that dependency depth admits an encoder-enforced bound---and we report each
measured limit together with the mechanism that sets it. Seventeen rejected
directions are listed with their numbers, including one that improved
density by 26\% and was declined. Every claim carries a level: reproducible by
command, measured with a stated reason, or estimated. The measurement tool is
released separately (DOI 10.5281/zenodo.22713364) with 435 provenanced records.
\end{abstract}


\section{Introduction}
The specifications of zlib, Brotli, LZ4 and Zstandard contain the same
sentence: the format does not attempt to provide random access to compressed
data. It is therefore bolted on afterwards---by our count in at least nine
independent implementations, including zstd's own seekable format, zeekstd,
seekable-zstd, two Go ports, one for .NET, \texttt{SeekableXZInputStream},
Google's RAC, and the \texttt{.zsi} indices from Clonezilla.

The trade this creates has been studied since 2016. Moffat and
co-authors~\cite{moffat} built the frontier for six methods across 36 points at
16, 64 and 256\,KiB. RAGE~\cite{rage} states it as common ground: the
granularity of random access is set by the block size. The xz documentation
says it plainly---smaller blocks mean faster seeks and worse compression. A
2026 challenge over 117 compressors~\cite{ait} reaches the same conclusion by
another route: no compressor dominates on every criterion.

Genomics pays this price knowingly rather than by accident: CRAM~3.1 caps
FQZComp at sixteen bits of context so that the model can be fitted quickly,
which is what makes it usable in a format built for random
access~\cite{cram}. What has not been done is to put a number on it.

\textbf{We do not discover the trade-off. We measure its steepness, and show a
format where the curve is nearly flat.} What follows from flatness is the
subject of the paper: if independence is cheap, a set of properties that are
normally mutually exclusive can be held at once.

\section{Format and What Follows}

\subsection{The core}
Every back-reference is resolved to an absolute position in the decompressed
output at encode time. Four separate streams are stored---literals, offsets,
lengths, commands---and the block table holds prefix sums into the decompressed
streams. Independence comes from exactly one restriction, that match search may
not cross a block boundary (\texttt{cur >= bstart}); the entropy streams remain
global. The specification is \texttt{docs/FORMAT\_STREAMS.md} in the public
repository.

\subsection{The cost of independence}
Two definitions are in circulation and both are legitimate; the paper must say
which it uses. We publish both, because for a month we published payload
figures without naming the convention, and an external check found it.

\emph{Payload} counts the compressed streams only, excluding the header and the
block table: chr1 0.41\%, enwik9 2.54\%, FASTQ 1.93\%. \emph{Archive} counts the
file on disk. Fine-grained index structures are usually dismissed on the grounds that the
index files become impractical~\cite{lcqs}; ours is measured rather than
assumed. On chr1 the archive is 80{,}829{,}622 bytes at 16\,KiB against
79{,}510{,}864 as a single block. The difference of 1{,}318{,}758 bytes is
1.632\% of the archive, of which the block table---64 bytes per block over
15{,}499 blocks, or 991{,}936 bytes---accounts for 75.2\%. All archive figures
below use the archive convention.

Table~\ref{tab:cg} gives the curve at five granularities, measured by an
independent agent on a different machine.

\begin{table}[t]
\caption{Cost of independence by granularity, archive convention, chr1;
measured by an independent agent on a Xeon 6973P-C.}
\label{tab:cg}
\centering
\small
\begin{tabular}{lrr}
\toprule
granularity & zstd-seekable & ACEAPEX \\
\midrule
4\,KiB   & 10.06\% & 5.33\% \\
16\,KiB  & 6.57\%  & 1.632\% \\
64\,KiB  & 2.83\%  & 0.598\% \\
256\,KiB & 4.30\%  & 0.267\% \\
1\,MiB   & $-0.51$\% & 0.150\% \\
\bottomrule
\end{tabular}
\end{table}

\begin{figure}[t]
\centering
\includegraphics[width=\columnwidth]{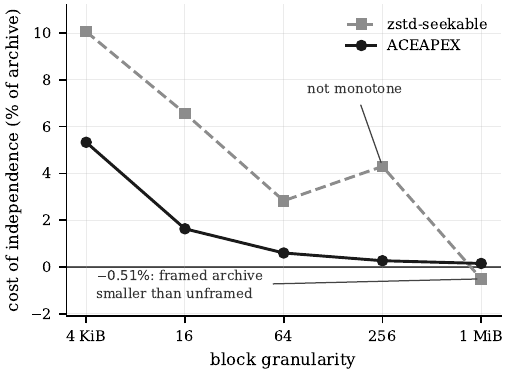}
\caption{The cost of independence falls with granularity three to four times
faster for one format than for the other; the zstd curve also bends upward at
256\,KiB and goes negative at 1\,MiB. Archive convention, chr1.}
\label{fig:cg}
\end{figure}

Two features of the zstd column deserve saying rather than smoothing: it is
\emph{not monotone}---256\,KiB is worse than 64\,KiB---and its last point is
negative. Both were re-run separately and every archive size matched byte for
byte, on the same corpus with a pinned codec version.

For bgzip no comparable single-parameter baseline exists, since gzip's 32\,KiB
window makes the base already blockwise. That cell is \texttt{n/a} with a
reason, not an omission.

\subsection{Cutting an archive is free}
Because blocks are independent and the entropy chunks follow the data, dividing
an archive costs nothing and can pay. Splitting chr1 in half on a block-aligned
boundary gives 80{,}663{,}353 bytes against 80{,}829{,}622 whole, $-0.206\%$;
tiling a 64\,MB prefix into sixteen 4\,MB pieces gives 22{,}966{,}121 against
23{,}030{,}417, $-0.279\%$.

Each part gets entropy chunks fitted to its own statistics. We state the
mechanism as a hypothesis rather than a finding: zstd also updates its tables
within a frame, so ``local statistics'' does not explain the whole effect.

The practical consequence is that a container of sub-archives gives append-only
behaviour with no format change. The overhead is $68 + 64b$ bytes for $b$
blocks: 3.3\% at one block, 1.3\% at 64.

\subsection{Latency as a parameter}
Table~\ref{tab:profiles} lists four configurations. None was chosen by hand;
all four are points on a measured Pareto frontier drawn from a sweep of 324
configurations, every one bit-perfect.

\begin{table}[t]
\caption{Profiles as points on a measured frontier. Payload convention,
EPYC 4344P.}
\label{tab:profiles}
\centering
\small
\begin{tabular}{lrrrrr}
\toprule
profile & block & LIT & FSE & seek p50 & ratio \\
\midrule
interactive & 16\,KiB & 64\,KiB & 4\,KiB & 0.084\,ms & 3.655 \\
seekable & 256\,KiB & 256\,KiB & 32\,KiB & 0.49\,ms & 3.769 \\
dense & 256\,KiB & 1\,MiB & 32\,KiB & 0.85\,ms & 3.778 \\
fast-decode & 256\,KiB & 1\,MiB & 4\,KiB & --- & 3.761 \\
\bottomrule
\end{tabular}
\end{table}

\section{Three Results}

\subsection{The repeat-distance chain is a substitution monoid}
A cache of the last $k$ distances, updated by permutation and by insertion of
explicit constants, is closed under composition: each command defines a map
whose every output slot is either a constant or a copy of an input slot. The
class is closed, associative, and has an identity. Therefore all states are
computable in $O(kn)$ work and $O(k\log n)$ span \emph{without altering the
bitstream}; for $k=4$ that is $O(n)$ work and $O(\log n)$ span.

Verification is per command over 21{,}132{,}882 commands across two opposing
profiles, with zero discrepancies.

The physical result depends on how loaded the device is, and we report the
dependence rather than a single number. On an H100: $2.87\times$ at 8192
blocks, $1.81\times$ at 15{,}499, $1.82\times$ at 31{,}000. On an RTX PRO 4000:
$14.8\times$ at 256 blocks, $2.1\times$ at 4096, $1.12\times$ at 8192, and
$0.87\times$ at 15{,}499. The crossing point sits between 8192 and 15{,}499
blocks on the weaker card and beyond our data on the H100. The mechanism is
that the sequential kernel is latency-bound and plateaus while the scan kernel
is throughput-bound and grows, so two lines cross. Small sets overstate the
gain badly---256 blocks give $38\times$, which measures kernel launch rather
than work---so the publishable figure is the saturated one.

\begin{figure}[t]
\centering
\includegraphics[width=\columnwidth]{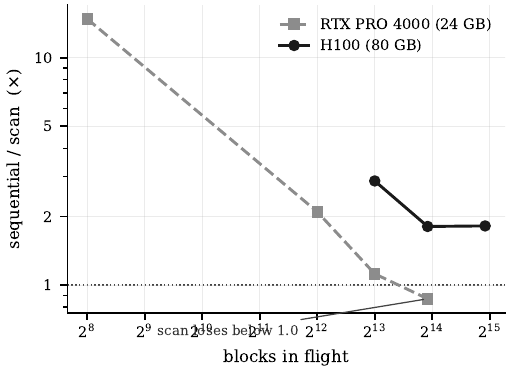}
\caption{The scan's advantage is a function of how loaded the device is, not a
property of the algorithm: on the weaker card it falls below the sequential
kernel once enough blocks are in flight, and on the H100 that crossing lies
beyond our data.}
\label{fig:scan}
\end{figure}

We are explicit about novelty. Scanning by function composition is old:
Ladner and Fischer~\cite{ladner} in 1980, and standard for finite automata.
What is new is that the repeat-distance parse of LZ77, which the field treats
as sequential by construction, falls into that class, and that the gain costs
no ratio.

\subsection{Self-overlap is a period, not a chain}
When $dist < len$ a decoder is obliged to copy byte by byte, each output byte
depending on the one just written. This is the sequential dependency behind the
belief that LZ77 cannot be parallelised.

The overlapping region is periodic with period $dist$, so
$\texttt{out}[dst+k] = \texttt{out}[src + k \bmod dist]$, and the source lies
entirely \emph{outside} the range being written. Every $k$ is independent and a
warp writes 32 bytes at once. Measured across four corpora, the match layer
speeds up by $2.75$--$8.42\times$, bit-perfect; this was first published in
Paper~5 of this series~\cite{aceapex}, and we restate it here because it is one
of the three places where the sequential reading of the format turns out to be
wrong.

The periodicity itself is neither new nor ours. Production CPU decoders use it
already: zstd's \texttt{ZSTD\_overlapCopy8} carries the tables
\texttt{dec32table = \{0,1,2,1,4,4,4,4\}} and \texttt{dec64table}, LZ4 carries
the same, and an open pull request vectorises them through \texttt{pshufb} and
NEON \texttt{tbl}. What differs is the width. There the period is used to
straighten the source so that eight bytes can be copied at once when
$\textit{offset} < 8$; here it is the width of a warp, because every offset
inside the match is computed independently rather than walked.

\subsection{An encoder-enforced bound on depth}
$1 + \max(\textit{level}_{\textit{src}}) \le L \Rightarrow \textit{MaxLevel}
\le L$, by induction, bit-perfect over the whole of chr1. The ratio cost runs
from zero to 0.0053\%, and at $L{=}32$ compression slightly improves.

A separate lever targets latency rather than structure. The MAX cap bounds the
maximum and costs nothing but does not touch latency; the AVG cap forces
literals where $\textit{lev}>1$ inside blocks whose average exceeds 2. It moves
the spike cluster's average from 4.36 to 0.56 and P99 from 4.33 to 1.50, for
0.081\% of ratio---sixteen times cheaper than raising the minimum match
length---with 1.223\% forced literals, bit-perfect.

The distinction matters because it closed a conceptual gap in the previous
paper: the spike cluster has eight times the median average depth, yet its
maximum stays below $L{=}32$, so a MAX cap does not see it at all.

\section{Limits, Each With Its Mechanism}
Table~\ref{tab:limits} lists what we hit and why.

\begin{table}[t]
\caption{Measured limits and their mechanisms.}
\label{tab:limits}
\centering
\small
\begin{tabular}{@{}lll@{}}
\toprule
axis & limit & mechanism \\
\midrule
GPU decode & 179.8\,GB/s & work granularity, median match 7\,B \\
CPU decode & 2564\,MB/s & Amdahl ceiling $1.75\times$, literals 43\% \\
parse & 64--72\% & three sequential parts, one removed \\
GPU seek & 360\,\textmu s & 282 of it fixed launch cost \\
encode & 232\,MB/s & LZ77 phase 91\%, 224 instr/byte \\
bus efficiency & 4.4\% & a property of the data \\
\bottomrule
\end{tabular}
\end{table}

\subsection{Block size, an axis nobody swept}
On an H100 the format reaches 179.8\,GB/s at 4\,KiB, 160.7 at 8\,KiB and 128.0
at 16\,KiB; on an RTX PRO 4000, 112.8 at 8\,KiB, 99.2 at 16\,KiB and 78.8 at
32\,KiB. The direction is the same on both cards---smaller blocks mean more
independent units of work and higher occupancy---and 179.8 exceeds the 172
published earlier in this series.

Ratio barely moves across an eightfold change: 3.174, 3.178, 3.181. The archive
behaves differently, because the block table doubles: 2.42\% at 8\,KiB against
1.23\% at 16\,KiB.

\begin{figure}[t]
\centering
\includegraphics[width=\columnwidth]{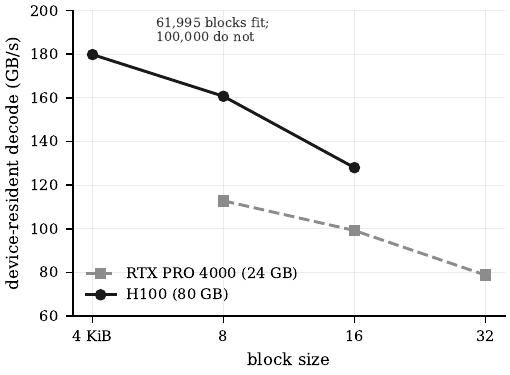}
\caption{Smaller blocks mean more independent units of work and higher
occupancy; the direction is the same on both cards, and the usable left end is
set by how many blocks fit in VRAM.}
\label{fig:bs}
\end{figure}

The VRAM limit is set by the \emph{number of blocks}, not by volume: 80\,GB
holds 61{,}995 blocks and fails at 100{,}000; 24\,GB holds 30{,}997. The
optimal block size therefore depends on file size, and the
\texttt{ACEAPEX\_BS >= 4096} floor in the code is physics rather than caution:
1 and 2\,KiB do not fit even on 80\,GB.

\subsection{On the CPU there is no memory wall}
Holding match length fixed at a median of 5 and varying only distance gives
13.8\,ns per match when everything is in L1, 12.5 on real chr1, 12.6 beyond
L1d, 12.8 beyond L2 and 13.8 beyond L3. There is no difference. The cost is
instructions, not memory: with a median distance of 81 bytes the source was
written a few cycles ago and is always in L1.

The wall exists on the GPU and only on the \emph{write} side. That is the 4.4\%
bus efficiency---store coalescing at a median length of 5.9 bytes.

\subsection{Match statistics}
From 1{,}508{,}720 matches over 4000 blocks of chr1, with the command stream
decoded against the encoder and coverage summing to exactly 100\%: length has
median 7 and mean 8.03, but the shape matters more than the summary. Almost
half of all matches---45.33\%---are exactly the minimum length of six bytes,
and the distribution is not unimodal: 15.04\% at seven, a second rise to
18.51\% at eight, then 7.33\% and 3.70\%, 7.42\% over 11--16 and 2.62\% above
16. Ninety per cent of matches are eleven bytes or shorter.

Matches cover \textbf{18.5\%} of the decompressed output of chr1 and literals
cover 81.5\%, summing to 100\%. That single pair explains several things at
once: why a domain transform on literals is worth 17\% of density, why the
median match is short, and why the match layer works on a fifth of the data
while the rest of the density is entropy. Distance has median
81 and p10 8, with 17.23\% at $d\le16$, 27.37\% at $\le32$, 43.83\% at
$\le64$, 66.64\% at $\le128$, 72.42\% at $\le256$, 79.19\% at $\le512$,
85.71\% at $\le1024$, 92.23\% at $\le2048$, 96.99\% at $\le4096$, 99.35\% at
$\le8192$ and 100\% at the block size of 16{,}384. Destination alignment is a multiple of 32 in 3.09\% of cases against
a random expectation of 3.12\%---that is, not at all.

\begin{figure*}[t]
\centering
\includegraphics[width=\textwidth]{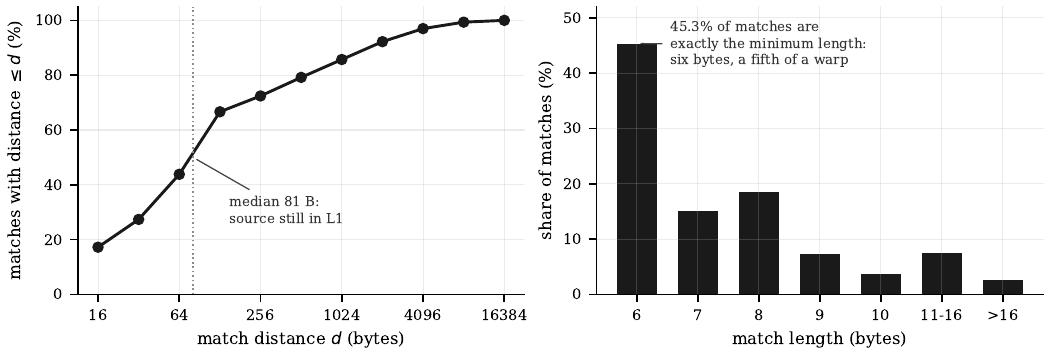}
\caption{Two distributions explain two of the limits: sources sit close enough
that they are still in L1 when reused, which is why there is no memory wall on
the CPU; and matches are shorter than a warp, which is why the GPU write bus
runs at 4.4\%. Both panels come from the same parse of $n = 1{,}508{,}720$ matches over 4000
blocks of chr1.}
\label{fig:dist}
\end{figure*}

\section{The Trade-off Measured, Nine Axes}
One operation is timed: a 16{,}000-byte logical region at a random uncompressed
offset, 200 requests, archive resident, timer around the API call only.
Table~\ref{tab:nine} gives four format-profile pairs.

\begin{table}[t]
\caption{Region read, 16\,kB logical, 200 requests. Archive convention,
Xeon 6973P-C (independent agent).}
\label{tab:nine}
\centering
\small
\begin{tabular}{@{}lrrrrr@{}}
\toprule
format / profile & ratio & p50\,ms & p99\,ms & amp. & b/even \\
\midrule
bgzip + htslib & 3.3826 & 0.0955 & 0.2048 & 4.82$\times$ & 3699 \\
zstd-seekable & 3.0258 & 0.0470 & 0.0545 & 2.00$\times$ & 7446 \\
ACEAPEX inter. & 3.6585 & 0.1308 & 0.2340 & 6.22$\times$ & 1386 \\
ACEAPEX dense & 3.7806 & 1.2805 & 2.4660 & 83.4$\times$ & 131 \\
\bottomrule
\end{tabular}
\end{table}

Throughput on an EPYC 7763: bgzip encodes at 19.8\,MB/s and decodes at 582.5;
zstd-seekable 150.4 and 482.3; ACEAPEX 39.5 and 1281.0. The dense profile
encodes at 39.2 and its full-decode figure is a data edge---three points
diverged by more than 5\%---so we do not publish it.

\begin{figure}[t]
\centering
\includegraphics[width=\columnwidth]{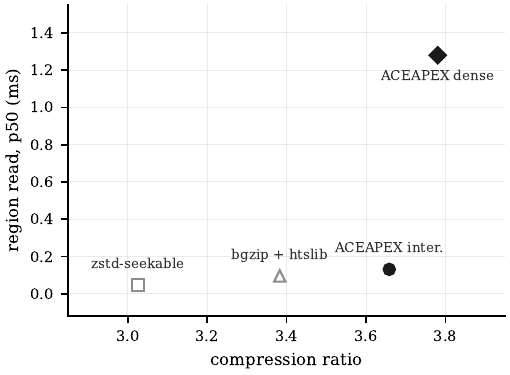}
\caption{No format dominates: the two established ones read a region faster,
ours are denser, and the dense profile trades an order of magnitude of latency
for the last 3\% of ratio. Xeon 6973P-C, chr1, 16\,kB logical region.}
\label{fig:frontier}
\end{figure}

\textbf{The losses belong in the text, not in a footnote.} ACEAPEX encodes
$3.81\times$ slower than zstd. zstd-seekable reads a single region three times
faster than we do and reads half as many bytes doing it. What we win is
density, above both, and full decode $2.20\times$ faster than bgzip and
$2.66\times$ faster than zstd-seekable.

\subsection{A latency law, and an open question inside it}
Across nine chunk configurations with amplification from $4.75\times$ to
$40\times$, region latency fits
$0.056\,\text{ms} + 0.0083\,\text{ms}\times\text{amplification}$ with
$R^2 = 0.93$, where amplification is
$(\textit{LIT}+3\,\textit{FSE})/\textit{block}$.

The intercept of 56\,\textmu s is 69\% of the fastest point and does not move
with chunk size; four zstd frames are touched per region.
\textbf{This is an open question of the paper, not a result.}

The fit is drawn from a chunk sweep on one machine (EPYC 4344P); the profiles
of Table~\ref{tab:nine} were measured elsewhere (Xeon 6973P-C) and are not
expected to lie on it. The intercept belongs to the first machine, and on the
second it differs.

\subsection{Batching, and a claim we had to correct}
Requests are grouped by block before threads start, with a threshold of 512
below which we go sequential. A previously circulated figure of $13\times$
compared eight threads against one. At equal thread count the gain is
$1.84\times$ (interactive: 5944 against 10912). A claim of this kind is
obliged to name the thread count.

The gain depends on request count---$1.05\times$ at 100, $2.75\times$ at 600,
$3.77\times$ at 2000, $4.60\times$ at 5000---and on the entropy $H_\alpha$ of
the access distribution over blocks: $4.6\times$ at 11.8 bits, $13.4\times$ at
6.3 (Zipf 1.2), $48.6\times$ at 5.8 (hot set). 244{,}000 requests were checked
byte for byte.

\subsection{Where random access stops paying}
At roughly \textbf{700 requests} on this file, full decode becomes cheaper---
that figure is from the EPYC 4344P, taken where the batch and full-decode
curves cross. The tool's own break-even for the same profile on the Xeon is
1386. Both are cost models on their own machine, and we give both rather than
choosing. Across the four formats the tool reports 3699 for bgzip, 7446 for
zstd-seekable, 1386 for the interactive profile and 131 for dense. We are not
aware of anyone publishing these numbers.

\section{One Archive, Two Consumers}
The full device-resident pipeline runs entropy and match on the GPU, bit-perfect
with FNV matching the original. An H100 80\,GB reaches 179.8\,GB/s at a 4\,KiB
block with $G{=}16$; an RTX PRO 4000 Blackwell 24\,GB reaches 112.8 at 8\,KiB,
both on driver 580.178.04 under CUDA 12.4.131.

Three H100s give 171.9, 171.5 and 171.5\,GB/s simultaneously, all bit-perfect---a
direct consequence of block independence: the file divides into $N$ parts and
each decodes without knowing about the others.

$G{=}16$ is confirmed as the default: at a 4\,K block, $G{=}8$ gives 156.4,
$G{=}16$ gives 179.3 and $G{=}32$ gives 171.7. The single exception is a 16\,K
block on chr1, where $G{=}32$ gives 136.2 against 129.7.

\textbf{GPU seek loses to CPU seek on both cards}: 360\,\textmu s on the H100
and 575--783\,\textmu s on the RTX PRO 4000, a 36\% spread, against 82\,\textmu s
on the CPU. The GPU is for streaming and for batches, not for a single read.

\texttt{ncu} was unavailable three times (\texttt{ERR\_NVGPUCTRPERM} in
containers without \texttt{--privileged}), so occupancy is argued indirectly
through the block-size sweep rather than from counters.

\section{What We Rejected}
The acceptance rule is that nothing already achieved may get worse; a trade is
declined regardless of how the arithmetic of the gain looks.

The hardest case was order-1 rANS on literals: 2.113 bits per byte against
zstd's 2.861, a 26\% density gain taking ratio from 3.181 to 4.105---and decode
fell 11\% with seek down 43\%, so it was declined. Checked against
Turbo-Range-Coder, whose adaptive order-1 gives 136\,MB/s at zstd's density
where ours gives 728\,MB/s at density 26\% better. The decision held.

Sixteen others, each with its number: a 6-byte hash gains 16\% speed and loses
1.7\% ratio on every corpus; a direct 8-mer table is four times faster than the
hash and finds half the matches; parsing for write geometry has nothing to
align, since destinations are effectively random (3.09\% against 3.12\%) and
matches longer than 32 bytes are 0.07\% of tokens and 0.5\% of data; SIMD search
without tables is 7--16 times slower and finds half as much; aggregating short
matches fails because the median run of consecutive short matches is 1; a 2-bit
decode window fails because $d$ is a multiple of 4 in 24.02\% of cases against a
random 25\%; a 64-byte register window is $1.61\times$ slower because
maintaining it costs more than it saves; transposing FASTQ loses 709\% on bases,
since correlation runs along a read and not across; order-1 on the command
stream gains 6.3\% over an alphabet of 249, which is 0.44\,MB for a second
coder; deriving the block table saves 0.70\% of size for 4.6\% of time; merging
the DNA streams costs 1.3\%; and skipping an empty length stream gains nothing.

Four directions were closed earlier and belong in the same list, since a closed
direction is also a result: a modular layout of literals (\emph{mod-4}) cost
0.9\% of ratio and gave nothing back at decode; a flat literal buffer in place
of chunks cost 43\% of seek; a 4-byte stride on DNA failed for the same reason
transposition does, correlation running along a read rather than across; and
separate decoders per stream (\emph{heterodecode}) added 11\% of time for no
change in ratio. Two further items on early lists---generation indexing and the
dual-stream bitstream---are not rejected ideas but features already implemented.

That is seventeen rejected directions in total.

\section{Method Is a Contribution}
\textbf{The absolute belongs to the machine; the ratio is what is checked.} One
regional call takes 0.082\,ms on an EPYC 4344P, 0.111 on a laptop under WSL,
0.131 on a Xeon 6973P-C and 0.154 on an EPYC 9V74---a factor of two---while the
ratio against bgzip stays between 10.8 and 15.2.

\textbf{Saturation.} Measuring on the whole of an available file measures the
corpus, not the system. Load must grow until the curve flattens; otherwise the
figure is a \emph{data edge}. We twice recorded an understated GPU number, 94
and then 136, because we were measuring the edge.

\textbf{Configuration is part of the name of a row}, written as
\texttt{ACEAPEX @ SHA · block 16\,KiB · lit 64\,KiB · fse 4\,KiB · threads 8}.
Library version is printed beside the number, since libzstd 1.5.5 is about a
percent denser than 1.4.8.

\textbf{A contract of 32 checks} runs from a clean clone with one command,
downloads chr1 from UCSC and verifies its md5, and runs in CI on every commit.
Claims carry levels: \textbf{R} reproducible by command with an expectation and
a tolerance, \textbf{M} measured with the reason it is not reproducible, and
\textbf{E} estimated from measured quantities.

\textbf{External checking finds what the home machine cannot.} In one day,
external sources---CI, a laptop under WSL, an independent agent---found five
defects and the main machine found none: a reused variable meant the report had
not been written since 13 August; a threshold taken from one machine did not
hold on three others; a $13\times$ batch figure was comparing eight threads
against one; a ratio was computed without the header and block table; and a read
past the end of an array when the block equalled the file size (found by ASan).
A sixth came through someone else's measurement: the encoder default was losing
17.1\% of density on genome, because literal chunking was off without an
explicit \texttt{LIT\_CHUNK} and the domain transform did not run without it
(3.18065 against 3.72329).

This is a methodological claim of the paper: a machine where everything is
configured cannot verify reproducibility---it is the thing one needs to be
independent of.

\section{The Tool}
\texttt{hw-apex-bench} is released separately under Apache-2.0 with measurements
under CC BY 4.0, DOI \texttt{10.5281/zenodo.22713364}: four codecs, nine axes,
435 records with provenance. A codec is added as a single adapter file;
\texttt{--check} builds it, runs a round trip and prints the axes it supports,
and \textbf{an unchecked adapter is not admitted to measurement}---which neither
lzbench nor TurboBench enforces. An unsupported axis prints \texttt{n/a} with a
reason supplied by the adapter, never an empty cell. The interface has been
exercised by an external CLI adapter with no change to the core.

\section{Limitations}
Encode is $3.81\times$ slower than zstd, because match search is confined to a
block and the table does not amortise across boundaries. This is the same stick
as the 1.632\% of Section~2---one cost seen from two ends.

A single region read in zstd-seekable is three times faster than ours at half
the amplification.

SAGe states our problem in the same words---genomic data is held compressed
and must be decompressed and reformatted before an accelerator can touch
it---and solves it in hardware, inside the storage device~\cite{sage}. We solve
it in the format, by making the compressed bytes directly readable. The two
approaches are complementary and neither subsumes the other.

A third line declines to change anything at all. Seekable OCI~\cite{soci}
traces \texttt{zran.c} through stargz to AWS SOCI and the zran mode of Nydus,
and concludes that no new format is needed---only an external index mapping
files to compressed byte ranges inside the existing gzip layers, which works
because a DEFLATE block decompresses independently given a starting offset and
32\,KiB of dictionary state. That is the price, and it is worth stating beside
ours: 32\,KiB of state per entry point there, 1.632\% inside the format here
and no external index. These are production systems at large scale solving the
same problem by the opposite route, and both work.

Our guarantee is also of a different kind from the theoretical one. LZBE
supports $O(\log n)$-time random access to a symbol~\cite{lzbe}, and lower
bounds for grammar-compressed strings~\cite{cicalese,gclower,kempa} say where
the ceiling of that line lies. We offer no asymptotic bound per symbol; we offer
block independence by construction and measured microseconds for a 16\,KiB
region. Neither result implies the other, and a reader from either side should
know which they are getting. Highly repetitive collections are a weakness: a match cannot
leave its block, so repeats between versions of a file are not taken, and
r-index style structures win there. The GPU figures come from rented cards
rather than CI, since GPU runners for open projects effectively do not exist.
The $c(g)$ curves rest on one corpus; other data will move them, which is part
of why the tool exists.

\textbf{There is no second independent decoder.} The specification is published
and nobody has written to it. There is no conformance corpus: there has been
fuzzing, but no corpus.

Of five gates toward a standard, two and a half are passed: the specification is
published, an implementation works, an external independent benchmark exists,
and there is one external user. A second independent decoder and a conformance
corpus are not---and the first of those is the one thing that cannot be done
alone. The code is in lzbench (PR \#276, \#277) under the maintenance of inikep
and tansy.

\section*{Reproducibility}
Canonical corpus: chr1, UCSC hg38, 253{,}935{,}557 bytes, md5
\texttt{9465e0f0df6e2c6eb39729c39cee5465}. Release tag \texttt{paper6-v1},
commit \texttt{b9e29102477f82a498ba49d15dc49306fb8c0013}. The ACEAPEX artifact
is archived at DOI \texttt{10.5281/zenodo.22758786} for this version, with
\texttt{10.5281/zenodo.20440964} resolving to all versions; the measurement
tool is at \texttt{10.5281/zenodo.22713364}. ACEAPEX is MIT licensed; the tool
is Apache-2.0 with its measurements under CC BY 4.0.

The contract runs from a clean clone with

\begin{quote}\small
\texttt{git clone https://github.com/yasha1971-coder/aceapex}\\
\texttt{cd aceapex \&\& ./reproduce\_paper5.sh}
\end{quote}

\noindent and records 32 pass, 0 fail, 9 skipped on an EPYC 4344P, and 19 pass,
0 fail, 17 skipped on a four-core GitHub Actions runner. The difference is the
GPU and large-corpus claims, which that runner cannot reach.

The parametric sweep of 324 configurations from which the four profiles of
Table~\ref{tab:profiles} are drawn is published as \texttt{sweep.jsonl} in the
same repository, together with \texttt{sweep\_all.sh} that produced it.


\begin{thebibliography}{99}
\bibitem{moffat} A.~Moffat \emph{et al.}, ``Access time tradeoffs in archive
compression,'' arXiv:1602.08829, 2016.

\bibitem{rage} C.~D.~Rask and D.~E.~Lucani, ``RAGE for the machine: image
compression with low-cost random access for embedded applications,''
arXiv:2402.05974, 2024.

\bibitem{ait} A.~Ribeiro \emph{et al.}, ``The 2026 algorithmic information
theory data compression challenge,'' arXiv:2606.17712, 2026.

\bibitem{ladner} R.~E.~Ladner and M.~J.~Fischer, ``Parallel prefix
computation,'' \emph{J.~ACM}, vol.~27, no.~4, pp.~831--838, 1980.

\bibitem{lzend} I.~Boneh and P.~Gawrychowski, ``Random access to LZ-End,''
arXiv:2607.14923.

\bibitem{lzbe} H.~Shibata, Y.~Nakashima, Y.~Yamaguchi, and S.~Inenaga,
``LZBE: an LZ-style compressor supporting $O(\log n)$-time random access,''
arXiv:2506.20107v3, 2026.

\bibitem{cicalese} F.~Cicalese, T.~Gagie, Z.~Lipt\'ak, G.~Navarro, N.~Prezza,
and C.~Urbina, ``Incongruity-sensitive access to highly compressed strings,''
in \emph{Proc.\ 34th European Symposium on Algorithms (ESA)}, LIPIcs vol.~388,
pp.~125:1--125:22, 2026. doi:10.4230/LIPIcs.ESA.2026.125.

\bibitem{gclower} A.~Duyster and T.~Kociumaka, ``Random access in
grammar-compressed strings: optimal trade-offs in almost all parameter
regimes,'' arXiv:2602.10864v2, 2026.

\bibitem{kempa} D.~Kempa and T.~Kociumaka, ``Tight lower bounds for central
string queries in compressed space,'' SODA, 2026.

\bibitem{sitaridi} E.~Sitaridi, R.~Mueller, T.~Kaldewey, G.~Lohman, and
K.~A.~Ross, ``Massively-parallel lossless data decompression,'' ICPP, 2016.

\bibitem{rapidgzip} M.~K\"ohler, T.~Bingmann, and P.~Sanders, ``Rapidgzip:
parallel decompression and seeking in gzip streams using cache prefetching,''
HPDC, 2023, pp.~295--307.

\bibitem{recoil} T.~Lin \emph{et al.}, ``Recoil: parallel rANS decoding with
decoder-adaptive scalability,'' ICS, 2023.

\bibitem{navarro} G.~Navarro, ``Indexing highly repetitive string
collections,'' \emph{ACM Computing Surveys}, 2021.

\bibitem{wesley} K.~Wesley, ``Massively parallel LZ77 compression and
decompression on the GPU,'' M.S. thesis, Dept.\ of Computer Science, Texas
State University, San Marcos, TX, Dec.\ 2022.

\bibitem{sage} N.~Mansouri~Ghiasi, T.~G\"uloglu, H.~Mustafa, C.~Firtina,
K.~Koliogeorgi, K.~Kanellopoulos, H.~Mao, R.~Nadig, M.~Sadrosadati, J.~Park,
and O.~Mutlu, ``SAGe: a lightweight algorithm-architecture co-design for
mitigating the data preparation bottleneck in large-scale genome sequence
analysis,'' in \emph{Proc.\ HPCA}, 2026. arXiv:2504.03732.

\bibitem{lcqs} J.~Fu, B.~Ke, and S.~Dong, ``LCQS: an efficient lossless
compression tool of quality scores with random access functionality,''
\emph{BMC Bioinformatics}, vol.~21, art.~109, 2020.
doi:10.1186/s12859-020-3428-7.

\bibitem{ffc} S.~Grabowski, T.~M.~Kowalski, and R.~Susik, ``FFC: a scalable
FASTA compressor,'' \emph{Bioinformatics}, vol.~42, no.~3, btag132, Mar.\ 2026.
doi:10.1093/bioinformatics/btag132.

\bibitem{dietgpu} Meta, ``DietGPU: GPU-based lossless compression,'' software,
\url{https://github.com/facebookresearch/dietgpu}.

\bibitem{nvcomp} NVIDIA, ``nvCOMP,'' software, repository archived 2026.

\bibitem{blackwell} A.~Jarmusch and S.~Chandrasekaran, ``Microbenchmarking
NVIDIA's Blackwell architecture: an in-depth architectural analysis,''
arXiv:2512.02189v3, 2026.

\bibitem{cram} J.~K.~Bonfield, ``CRAM 3.1: advances in the CRAM file format,''
\emph{Bioinformatics}, vol.~38, no.~6, pp.~1497--1503, Mar.\ 2022.
doi:10.1093/bioinformatics/btac010.

\bibitem{soci} J.~Thompson, W.~Mesard, J.~Butler, S.~S.~B.~Vellore~Rajakumar,
and H.~Wang, ``Seekable OCI: lazy-loading container images via range-request
indexing,'' arXiv:2607.06868, 2026.

\bibitem{aceapex} Y.~Shavidze, ACEAPEX Papers 1--5: arXiv:2606.04268,
2606.18900, 2606.24531, 2607.18541, 2608.10188.
\end{thebibliography}
\end{document}